\documentstyle{article}
\begin{document}
\def\nn{\nonumber \\}
\def\be{\begin{equation}}
\def\ee{\end{equation}}
\def\ba{\begin{eqnarray}}
\def\ea{\end{eqnarray}}
\def\la{\label}
\def\re{(\ref}

\def\i{{\rm i}}
\let\a=\alpha \let\b=\beta \let\g=\gamma \let\d=\delta
\let\e=\varepsilon \let\z=\zeta \let\h=\eta \let\th=\theta
\let\dh=\vartheta \let\k=\kappa \let\l=\lambda \let\m=\mu
\let\n=\nu \let\x=\xi \let\p=\pi \let\r=\rho \let\s=\sigma
\let\t=\tau \let\o=\omega \let\c=\chi \let\ps=\psi
\let\ph=\varphi \let\Ph=\phi \let\PH=\Phi \let\Ps=\Psi
\let\O=\Omega \let\S=\Sigma \let\P=\Pi \let\Th=\Theta
\let\L=\Lambda \let\G=\Gamma \let\D=\Delta

\def\0{\over } \def\1{\vec } \def\2{{1\over2}} \def\4{{1\over4}}
\def\5{\bar } 
\def\6{\partial }
\def\7#1{{#1}\llap{/}}
\def\8#1{{\textstyle{#1}}} \def\9#1{{\bf {#1}}}

\def\({\left(} \def\){\right)} \def\<{\langle } \def\>{\rangle }
\def\[{\left[} \def\]{\right]} \def\lb{\left\{} \def\rb{\right\}}
\let\lra=\leftrightarrow \let\LRA=\Leftrightarrow
\let\Ra=\Rightarrow \let\ra=\rightarrow
\def\ul{\underline}

\let\ap=\approx \let\eq=\equiv  
        \let\ti=\tilde \let\bl=\biggl \let\br=\biggr
\let\bi=\choose \let\at=\atop \let\mat=\pmatrix
\def\CL{{\cal L}} \def\CD{{\cal D}} \def\rd{{\rm d}} \def\rD{{\rm D}}
\def\CH{{\cal H}} \def\CT{{\cal T}}
\newcommand{\dR}{\mbox{{\it I \hspace{-0.9em} R}}}
\newcommand{\de}{\mbox{{\bf 1}}}

\newcount\nummer \nummer=0
\def\f#1{\global\advance\nummer by 1 \eqno{(\number\nummer)}
      \global\edef#1{(\number\nummer)}}

\begin{titlepage}
\renewcommand{\thefootnote}{\fnsymbol{footnote}}
\renewcommand{\baselinestretch}{1.3}
\hfill  TUW - 93 - 25 \\ \medskip \hfill hep-th/9401110 \\
\medskip
\vfill

\begin{center}
{\LARGE {Diffeomorphisms versus
Non Abelian Gauge Transformations:\\
An Example of 1+1 Dimensional Gravity}
 \\ \medskip  {}}
\medskip
\vfill

\renewcommand{\baselinestretch}{1} {\large {PETER
SCHALLER\footnote{e-mail: schaller@email.tuwien.ac.at} \\
\medskip THOMAS STROBL\footnote{e-mail:
tstrobl@email.tuwien.ac.at} \\ \medskip\medskip
\medskip \medskip
Institut f\"ur Theoretische Physik \\
Technische Universit\"at Wien\\
Wiedner Hauptstr. 8-10, A-1040 Vienna\\
Austria\\} }
\end{center}

\vfill
\begin{center}
submitted to: {\em Phys. Letts. B}
\end{center}
\vfill
\renewcommand{\baselinestretch}{1}                          

\begin{abstract}

We investigate the phase space of a typical model of
(1+1) dimensional gravity (Jackiw-Teitelboim model
with cylindrical topology) using its reformulation
as a non abelian gauge theory based on the $sl(2,\dR )$
algebra. Modifying the conventional approach we argue
that one should take
the universal covering of $SL(2,\dR )$
rather than $PSL(2,\dR)$ as the gauge group of the theory.
We discuss the consequences for the quantization of the
model and find that the spectrum of the Dirac observables
is sensible to this modification. Our analysis further provides
an example for a gravity theory where the standard Hamiltonian
formulation identifies gravitationally inequivalent solutions.
\end{abstract}

\vfill
\hfill Vienna, December 1993  \\
\end{titlepage}

The identification of diffeomorphisms and non abelian gauge
transformations is an important tool for the quantization of two
and three dimensional gravity theories \cite{Japetc}, \cite{Wit}.
Though this mechanism
seemingly does not work in four dimensions \cite{com}, one may gain from it
insight into the structure of quantum gravity theories, which
might be significant for the quantization of four dimensional
gravity, too. Thus it is worthwhile to investigate this
identification thoroughly. The aim of the present paper is to
perform such an investigation for the example of the (1+1) dimensional
Jackiw-Teitelboim model \cite{JT}.\footnote[1]{In a similar
analysis presented in
\cite{Span} precisely the two points explained in the following
text have been missed. This is due to subtleties in the special
parametrization of the metric chosen in \cite{Span} (e.g.\ the
two functions $A$ and $B$ in this parametrization may have poles
and still render a $C^\infty$-metric).}

In particular there are two points diserving a careful
analysis:\\[-1.5\baselineskip]
\begin{itemize}
\item[(1)] The identification of non abelian gauge transformations on the
one hand and gravitational symmetries
(diffeomorphisms and local Lorentz transformations)
on the other hand is defined on the Lie algebra level. It does
not necessarily hold for transformations not smoothly connected
to the unity. In (2+1) dimensional gravity it is well known that
big diffeomorphisms which cannot be identified with
gauge transformations play a crucial
role (e.g. modular transformations on the torus for a space time
manifold $T^2\times \dR$). In the case of
the (1+1)-dimensional Jackiw-Teitelboim model on a cylinder
the inverse effect
occurs:
Following the literature and taking $PSL(2,\dR)\sim SO_e(2,1)$ as the
gauge group of the model, one finds
gauge transformations not connected to the unity,
which cannot be identified with gravitational symmetries.
As we will show in the present article, this problem is
overcome replacing PSL(2,\dR) by
its universal covering \\[-1.5\baselineskip]
\item[(2)] Even at the infinitesimal level, the one to one
correspondence between gauge transformations and gravitational
symmetries breaks down for a degenerate space time metric
(i.e. $\det g=0$). At first sight one could
think this to be irrelevant, as $\det g=0$ corresponds to
unphysical solutions. This argument is, however, erroneous in
general. Let us illustrate this by the consideration of a simple
analogous situation. Take the real line and as a symmetry group
the group of translations with generator $T_1=\partial /\partial
q$, $q\in \dR$. The
application of this symmetry will reduce the real line to a
point. Take another group, whose infinitesimal generator $T_2$ is
identified with the generator of translations via $T_2=qT_1$.
Under the action of this group the real line will split into
three gauge orbits: $\dR^+$, $\dR^-$ and $\{0\}$. Even if the point
$q=0$, where the correspondence between the infinitesimal
actions of the two
groups breaks down, is eliminated, we end up with different
results. We will show in the present article that a similar
mechanism occurs in the case of (1+1) dimensional gravity:
Eliminating the solutions with $\det g=0$, the gauge orbits split
into components not smoothly connected to each other. Solutions
from different components of the same gauge orbit
are not related by gravitational symmetry
transformations. They correspond to space time manifolds with
different kink number \cite{kink}.\\[-1.5\baselineskip]
\end{itemize}

The action of the Jackiw-Teitelboim model, when written
in Cartan's formulation with zweibein $e^a$, $a \in \{+,-\}$,
spin-connection $\o$, and Lagrange multiplier fields $(B_a,B_2) $, reads
\be S = \int B_a(de^a + \e^a{}_b \o \wedge e^b) +
     B_2 (d\o + e^+ \wedge e^-) , \la{action}
\ee
The equations of motion yield a 1+1 dimensional space time with
constant curvature and vanishing torsion.
(The value of the curvature is proportional to a coupling
constant which has been fixed to one in (\ref{action})).
With the reinterpretation
\be A=e^a J_a + \o J_2, \qquad F=dA + A^2 \la{asd} \ee
where $J$ denotes the generators of the $sl(2,\dR)$ Lie algebra
\be [J_a,J_b] = \e_{ab} J_2, \qquad [J_a,J_2] = \e^b{}_a J_b, \la{alg}
\ee
S becomes the action of a $PSL(2,\dR )$-gauge theory
\be S = 2 \int tr BF . \la{ac2} \ee
The equations of
motion generated by (\ref{ac2}) yield the connection $A$ to be
flat ($F\equiv O$) and the Lagrange multiplier field $B$ to be covariantly
constant ($DB:=dB+[A,B]\equiv 0)$.

In addition to the identification of the actions (\ref{action}) and
(\ref{ac2}), (\ref{asd}) induces a relation between the symmetries of the
corresponding models:
The Lie derivative of a vector field $\xi$
acting on the one form $A$ can be written as
$L_{\xi}A = i_\xi F + D(i_\xi A)$
and therefore an infinitesimal diffeomorphism
parametrized by $\xi$
and an infinitesimal local Lorentz transformation parametrized by an angle
$\theta$ can be identified on shell with gauge transformations
$\delta A=D \Lambda$ via
$\Lambda = i_\xi A$ and $\Lambda = \theta J_2$, respectively.
It is easy to see that this identification extends to the
$B$-fields also.
With (\ref{asd}) the correspondence is seen to be one to one,
iff $\det e\neq 0$.

Up to gauge transformations a flat connection $A$
on a cylinder is determined
by its monodromy $M_A=P\exp\oint A \in PSL(2)$ generating
parallel transport
around the cylinder ($P$ denotes path ordering and the
integration runs over a closed curve $C$ winding around the
cylinder once). With a $2\pi$-periodic coordinate $x^1$
it can always be written as
$A=A_1dx^1$ where $A_1$ is constant and has one of the following
forms
\be
A^I=\left( \begin{array}{cc}  0 & \a \\
\a& 0 \end{array}\right)dx^1, \quad
A^{II}=\left( \begin{array}{cc}  0 & \Th\\
-\Th& 0 \end{array}\right)dx^1, \quad
A^{III}=\left( \begin{array}{cc}  0 &  0 \\
\pm 1 & 0 \end{array}\right)dx^1 \la{A} \ee
with $\a,\Th \in \dR $ and the identifications
$\a\sim -\a$ and $\Th \sim \Th + 1/2 $.
The three cases $I$, $II$, $III$ represent the hyperbolic, the
elliptic, and the nilpotent sector of the moduli space of flat
$PSL(2,\dR )$
connections on the cylinder, respectively. (Note that the nilpotent sector
consists of two points only. There seems to be some confusion
about that in the literature).
In all of these three cases B is constant. It has the form
\be
B^I=\left( \begin{array}{cc}  0 &c_1 \\
c_1& 0 \end{array}\right), \quad
B^{II}=\left( \begin{array}{cc}  0 & c_2\\
-c_2& 0 \end{array}\right), \quad
B^{III}=\left( \begin{array}{cc}  0 &  0 \\
c_3 & 0 \end{array}\right),
\qquad c_i \in \dR,  \la{mom} \ee
respectively.
In the case $\a =0$ we have the additional identification
$c_1\sim -c_1$ and the additional solutions
\be
B^{IV}=\left( \begin{array}{cc}  0 &  0 \\
\pm 1 & 0 \end{array}\right) . \la{B1} \ee
(\ref{A}-\ref{B1}) give a parametrization of the reduced phase space
of the $PSL(2,\dR)$-gauge theory. Concerning the topological structure
of the latter we want to mention here only that it is not Hausdorff.
A more detailed analysis may be found elsewhere.

Now, the group elements
 \be g_{(n)} = \left( \begin{array}{cc}
\cos (nx^1/2) & \sin (nx^1/2)\\
-\sin (nx^1/2) & \cos (nx^1/2) \end{array}\right)
\qquad n\in Z    \la{groupel}
 \ee
applied to $A^I$ and $B^I$ yield
\ba A^I_{(n)}&=& \left( \begin{array}{cc} \a\sin (nx^1)
 & \a\cos (nx^1) + n/2 \\
\a\cos (nx^1) - n/2 & -\a\sin (nx^1) \end{array}\right)dx^1
\nn
B^I_{(n)}&=&c_1\left( \begin{array}{cc} \sin (nx^1)  &  \cos (nx^1)  \\
 \cos (nx^1) & -\sin (nx^1) \end{array}\right) .\la{trsol}
 \ea
It is obvious that these solutions, equivalent in the
$PSL(2,\dR)$ gauge theory
formulatiom for fixed values of $\alpha$ and $c_1$ but
different values of $n$,
cannot be transformed into each other by
diffeomorphisms and local Lorentz transformations: The latter
cannot change the number of zeros of $B$, which depends on $n$.
This clearly indicates an inequivalence of
$PSL(2,\dR)$ gauge transformations and gravitational symmetry
transformations.

{}From a more general point of view the space of gauge transformations
is the group of smooth mappings $G=\{g:S^1\times \dR\to
PSL(2,\dR)\}$. There is a natural homomorphism from $G$
to the homotopy group $\Pi_1 (PSL(2,\dR)) $.
We thus have $\Pi _0(G)=\Pi_1 (PSL(2,\dR)) \sim Z $, i.e.
the space of gauge transformations consists of an
infinite number of
components not smoothly connected to each other.
The group of diffeomorphisms and local Lorentz transformations
consists of a finite number of such components only.
They differ by $x^0$- and $x^1$-reflection on the space time manifold
and by parity transformation and time
reversal in the Lorentz bundle.
Up to these transformations
the discrepancy
between gauge transformations and gravitational symmetry
transformations is removed by restricting the gauge
transformations to the component $G\vert _e$ of $G$ connected to
the identity in $G$ (and thus associated with the identity in
the homotopy group). This is equivalent to choosing the
universal covering $\widetilde {SL}(2,\dR)$ rather than
$PSL(2,\dR)$ as the gauge group of the theory: As
$\widetilde {SL}(2,\dR)$ is simply connected, the associated space of
gauge transformations $\widetilde G=\{g:S^1\times \dR\to
\widetilde{SL}(2,\dR)\}$ is connected.
We may see this equivalence also
from another point of view: The covering map induces a
natural identification $G\vert _e \sim \widetilde G /center(\widetilde
{SL}(2,\dR))$. The elements of the center, however, generate
trivial gauge transformations.
To further elucidate this abstract identification let us
choose a loop $C$ running around the cylinder
once and a point $\hat x \in C$.
The connection $A$ then generates a parallel transporter
$M(x)=P\exp \int_{\hat x}^{x\in C} A$ and thus a path in the gauge group
connecting the unity with the
monodromy matrix $M_A$ as defined above. The restriction
of a gauge transformation $g \in G$ to $C$ yields
a closed loop $\gamma$ in the gauge
group. $\gamma$ acts
on $M(x)$
according to $M(x) \to \gamma (x)M(x)\gamma (\hat x)^{-1} $.
We may split $\gamma$ into a constant group element and a based
loop $\hat \gamma $: $\gamma (x) = \gamma (\hat x)\hat\gamma (x)$,
$\hat\gamma (\hat x)=1$. $\hat\gamma$ identifies all paths $M (x)$
ending at the same monodromy matrix $M_A$ whereas $\gamma (\hat x)$ allows
to identify monodromy matrices related by the adjoint action of
the gauge group. With $PSL(2,\dR )$ being the gauge group, we end up
with the configuration space $PSL(2,\dR )/Ad_{PSL(2,\mbox{\scriptsize\dR} )}$.
A
restriction to gauge transformations connected to the unity
means to restrict $\hat\gamma$ to contractible loops.
The latter generate smooth deformations of $M(x)$ only and thus identify
paths corresponding to the same element in the universal
covering of the gauge group.
Thus, if we want to relate the BF theory to
the Jackiw-Teitelboim model we are led to consider
$\widetilde {SL}(2,\dR )/Ad_{PSL(2,\mbox{\scriptsize\dR} )}\sim
\widetilde {SL}(2,\dR )/Ad_{\widetilde {SL}(2,\mbox{\scriptsize\dR} )}$ rather
than
$PSL(2,\dR )/Ad_{PSL(2,\mbox{\scriptsize\dR} )}$ as the
configuration space of the theory. The extension of this considerations
to the entire phase space is straightforward.

A complete set of representatives for the
components of $G$ not smoothly connected to each other
is given by (\ref{groupel}).
Thus we obtain a complete set of gauge inequivalent solutions of
the gauge theory with gauge group $\widetilde {SL}(2,\dR )$ by
applying the gauge transformations (\ref{groupel}) to
(\ref{A}-\ref{B1}).
In the hyperbolic sector we obtain
(\ref{trsol}). In the elliptic sector the element $g_{(n)}$ induces
a shift $\Th \to \Th + n/2 $ in (\ref{A}). Thus the transition
to the universal covering simply removes the identification
$\Th \sim \Th + 1/2 $.

Analogous results are obtained in the remaining sectors
of the phase space.

Still, our analysis is not complete. With the identifications
(\ref{asd}) the solutions we obtain by
the action of (\ref{groupel}) on
(\ref{A}-\ref{B1}) correspond to space time manifolds with $\det g =0$.
To any of these solutions, however, it is possible to find
a gauge transformation yielding a solution corresponding to
a nondegenerate space time metric. More precisely, this can be done in
an infinite number of gravitationally inequivalent ways.
E.g. in the elliptic sector, we
might apply one of the following
gauge transformations to $A^{II}$
  \ba & g_{[k]} = \left( \begin{array}{cc}
    \cos \th_k & \sin \th_k \\
    -\sin \th_k & \cos \th_k \end{array}\right)
    \left( \begin{array}{cc} 1 & b_k \\ 0 & 1 \end{array}\right) & \nn
 & \!\! \th_k = [\exp (x^0)\,+2\vert\Th\vert ]\sin(kx^1), \,
b_k = [\exp(x^0)\,+2\vert\Th\vert ]\cos(kx^1)
   \quad k\in N .  &  \la{smallgrel}
\ea
We obtain
\be A^{II}_{[k]} = \left( \begin{array}{cc}
    b_k(\Th dx^1+d\th_k ) & (1+{b_k}^2) (\Th dx^1+d\th_k )+db_k \\
     (\Th dx^1+d\th_k ) & -b_k(\Th dx^1+d\th_k ) \end{array}\right)
     \la{novang} \ee
The gauge transformations (\ref{smallgrel})
are smoothly connected to the unity for arbitrary value of $k$
as the $\th_k$ are periodic functions in $x^1$. Nevertheless the solutions
$A^{II}_{[k]}$ are gravitationally inequivalent for different values of
$k$. To prove this let us again choose a loop $C$ running around the
cylinder once.
Under the restriction $\det g = \det e \neq 0$ the components of
the zweibein $(e_0^+, e_1^+)$ induce a map $C\sim
S^1\to \dR^2\backslash {\{0\}} $ characterized by a winding number
(not depending on the choice of $C$).
Solutions with different winding numbers
cannot be transformed
into each other by gravitational symmetries, since they are
separated by solutions with $\det e =0$. (Also the
discrete gravitational
symmetry transformations mentioned above do not change the
winding number).
For different values of $k$ the solutions (\ref{novang}) have
different winding numbers, which proves our assertion.

This result generalizes to the other sectors of the theory:
Solutions which are gauge equivalent in the
$\widetilde{SL}(2,\dR)$ gauge theory are not equivalent in the
gravity theory, if they have different winding number.

The winding number defined above is related to the kink number
as defined in \cite{kink} by means of 'turn arounds' of the light cone
along non contractible loops. More precisely, winding number $k$
corresponds to kink number $2k$. (Odd kink numbers \cite{kink}
characterize
solutions which are not time orientable. Such solutions are not
considered here).

The physical relevance of solutions with nontrivial winding number
is not quite clear. They necessarily contain closed lightlike curves.
There are, however, also solutions with trivial winding containing closed
lightlike curves. At this point it might be interesting to note that in a
conventional Hamiltonian treatment of the action (\ref{action}) the
constraints will generate infinitesimal gauge transformations rather than
gravitational symmetry transformations. Thus on the Hamiltonian level
the kink number will not appear in the parametrization of
the reduced phase space, while, however, not all solutions with closed
timelike curves can be excluded in this way. A similar
situation occurs also when treating other models of two
dimensional gravity (cf.\ e.g.\ \cite{Klo}). It would be
interesting to see, if the equivalence up to $\det g =0$
of the Hamiltonian and Lagrangian formulation of four
dimensional gravity leads to similarly inequivalent factoring
spaces.

Let us close this letter by some remarks on the quantization of the model:
The application of a Dirac quantization
procedure in a configuration space representation of the quantum theory
will yield a quantum Hilbert space consisting of functions over the space
$\widetilde {SL}(2,\dR )/Ad_{PSL(2,\mbox{\scriptsize\dR })}$ (rather than
${SL}(2,\dR )/Ad_{PSL(2,\mbox{\scriptsize\dR })}$ as suggested in
\cite{Japetc}).
The most
remarkable consequence of the transition to the universal covering
of the gauge group is that the
sector of the phase space corresponding to the
connections $A^I$ in (\ref{A}) (elliptic sector) is compact in
the BF theory, but becomes noncompact upon transition to the
universal covering of the gauge group. Thus, in this sector,
the corresponding momentum operator (i.e. the Dirac observable
$\det B$) will have a discrete spectrum in the $PSL(2,\dR )$-gauge
theory, but a continuous spectrum in the gravity theory.

\vskip7mm
One of the authors (T. S.) wants to thank M. Henneaux for
discussions at the $3^{\mbox{rd}}$ RIM Baltic Student Seminar, where
partial  results of this work have been presented (the
corresponding proceedings will be published in L.N.P.).

\end{document}